\documentclass[a4paper]{aa}
\usepackage{graphicx}
\begin{document}
\title{Formation of filament-like structures in the pulsar 
magnetosphere and the short-term variability of pulsar emission} 
\author{V.~Urpin\inst{1,2}}
\institute{$^{1)}$ INAF, Osservatorio Astrofisico di Catania,
           Via S.Sofia 78, 95123 Catania, Italy \\
           $^{2)}$ A.F.Ioffe Institute of Physics and Technology and
           Isaac Newton Institute of Chile, Branch in St. Petersburg,
           194021 St. Petersburg, Russia
}
\date{\today}

\abstract
{Magnetohydrodynamic (MHD) instabilities can play an important role
in the dynamics of the pulsar magnetosphere and can be responsible for 
the formation of various structures.
}
{We consider the instability caused by a gradient of the magnetic
pressure which can occur in a non-neutral magnetospheric plasma of 
the pulsars. 
} 
{Stability is discussed by means  of a linear analysis of the 
force-free  MHD equations.
} 
{We argue that the pulsar magnetospheres are always unstable. The
unstable disturbances have a form of filaments directed along the
magnetic field lines with plasma motions being almost parallel
(or anti-parallel) to the magnetic field. The growth rate of 
instability is high and can reach a fraction of $ck$,
where $k$ is the wavevector of unstable disturbances. The instability
can be responsible for fluctuations of plasma and the short-term
variability of pulsar emission.}
{}


\authorrunning{V.Urpin}

\titlerunning{Formation of structures in the pulsar magnetosphere}

\maketitle

\section{Introduction} 

The pulsar magnetospheres consist mainly of electron-positron 
plasma. This plasma can affect the radiation produced in the inner 
region of the magnetosphere or at the stellar surface and, owing to 
this, the pulsar emission can provide information regarding the 
physical conditions in the magnetosphere. This might be a powerful 
method for the diagnostics of the magnetosphere. For instance, fluctuations 
of the pulsar emission can be caused by non-stationary phenomena in the 
magnetospheric plasma (such as instabilities, waves, etc.), which are 
determined by the conditions in the magnetosphere. Therefore, the spectrum 
and characteristic timescale of the detected fluctuations can provide 
important information regarding the state of plasma in the magnetosphere. 
That is why understanding non-stationary properties of the magnetospheric 
plasma is of crucial importance for the interpretation of observational 
data. 

The physical processes in the pulsar magnetosphere are very particular. 
The mean free path of particles is typically short compared to the 
characteristic lengthscale and, hence, the magnetohydrodynamic discription 
is justified. For typical values of the magnetic field, the electromagnetic 
energy density is greater than the kinetic energy density. This suggests 
that the force-free equation is a good approximation for determining 
the magnetic field structure over much of the magnetosphere. The 
growing observational data on spectra and pulse profiles of isolated 
pulsars prompt continued improvement of theoretical models of such 
force-free magnetospheres (see, e.g., Goodwin et al. 2004, Contopoulos et 
al. 1999, Komissarov 2006, McKinney 2006, Petrova 2013). In the 
axisymmetric case, the equations governing the structure of a pulsar 
magnetosphere can be reduced to the well-known Grad-Shafranov equation 
(see, e.g., Michel 1973, Mestel 1973, Mestel \& Shibata 1994; see also 
Beskin 1997 for general overview)). Most models based on this 
equation have a ``dead zone'' with field lines that are close within 
the light-cylinder and a ``wind zone' with poloidal field lines 
that cross the light-cylinder. Poloidal currents in the ``wind zone'' 
maintain a toroidal field component, whereas currents are vanishing in 
the ``dead zone''.

Many magnetohydrodynamic (MHD) phenomena in the force-free magnetosphere, 
however, are still poorly understood. Particularly, this concerns the 
non-stationary processes, such as the various types of instability, 
that can occur in the magnetosphere. This question is of particular 
interest because the existence of a stationary force-free configuration 
even rises doubts (see, e.g., Timokhin 2006). One of the MHD instabilities that 
can arise in the pulsar magnetosphere is the so-called diocotron 
instability, which is the non-neutral plasma analog of the 
Kelvin-Helmholtz instability. This instability has been studied 
extensively in the context of laboratory plasma devices (see, e.g., 
Levy 1965; Davidson 1990; Davidson \& Felice 1998). The possible 
existence of pulsars having a differentially rotating equatorial 
disk with a non-vanishing charge density could trigger instability of 
diocotron modes (Petri et al. 2002). In the non-linear regime, the 
diocotron instability might cause diffusion outwardsof the charged particles 
across the magnetic field lines outwards (Petri et al. 2003). The 
role of a diocotron instability in causing drifting subpulses in 
radio pulsar emission has been discussed by Fung et al. (2006). 

Recently, a new mode of the magnetospheric oscillations has been 
considered by Urpin (2011). This mode is closely related to the 
Alfv\'enic waves from standard magnetohydrodynamics, which have
been modified by 
the force-free condition and non-vanishing electric charge density. 
This type of magnetospheric waves can be unstable because there is 
a number of destabilising factors in the magnetosphere (such as 
differential rotation, electric currents, non-zero charge density, 
etc.). For example, many models of the magnetosphere predict that 
rotation should be differential (see, e.g., Mestel \& Shibata 
1994; Contopoulos et al. 1999) but it is known that differential 
rotation in plasma with the magnetic field leads to the so-called 
magnetorotational instability (Velikhov 1959). In the axisymmetric 
model of a magnetosphere suggested by Countopoulos et al. (1999), 
the angular velocity decreases inversely proportional to the 
cylindrical radius beyond the light cylinder and even stronger in 
front of it. For such rotation, the growth time of unstable 
magnetospheric waves is of the order of the rotation period 
(Urpin 2012). Numerical modelling by Komissarov (2006) showed 
that plasma rotates differentially basically near the equator 
and poles within the light cylinder. Such strong differential 
rotation should lead to instability that also arises on a timescale 
of the order of a rotation period. Note that the magnetorotational 
instability in the pulsar magnetosphere differs essentially 
from the standard magnetorotational instability because of a 
non-vanishing charge density and the force-free condition 
(Urpin 2012). 

The electric currents flowing in plasma also provide a destabilising 
influence that leads to the so-called Tayler instability (see, 
e.g., Tayler 1973a, b). This instability is well studied in both 
laboratory and stellar conditions. It arises basically on the Alfv\'en 
time scale and is particularly efficient if the strengths of the toroidal 
and poloidal field components differ significantly (see, e.g., Bonanno 
\& Urpin 2008a,b). This condition is satisfied in many magnetospheric 
models (see, e.g., Contopoulos et al. 1999), and these models can be 
unstable. 

However, this instability might have a number of qualitative 
features in the pulsar magnetosphere because of the force-free condition 
and non-zero charge density. In the present paper, we consider the 
instability of the pulsar magnetosphere relevant to magnetospheric waves
considered by Urpin (2011). We show that these waves can be unstable
in the regions where the magnetic pressure gradient is non-vanishing.
The considered instability arises on a short timescale 
and can be responsible for a short-term variability of 
the pulsar emission.

\section{Basic equations}

Despite uncertainties in estimates of many parameters, plasma in 
the pulsar magnetosphere is likely collisional and the Coulomb mean 
free path of electrons and positrons is small compared to the 
characteristic length scale. Therefore, the magnetohydrodynamic 
description can be applied to such highly magnetized plasma (see
Urpin 2012 for more details). 

Let us define the hydrodynamic velocity and electric current of an
electron-positron plasma as
\begin{equation}
{\bf V} = \frac{1}{n} (n_e {\bf V}_e + n_p {\bf V}_p) , \;\;\;
{\bf j}= e (n_p {\bf V}_p - n_e {\bf V}_e),
\end{equation}
where $({\bf V}_e, n_e)$ and $({\bf V}_p. n_p)$ are the partial 
velocities and number densities of electrons and positrons, 
respectively; $n=n_e + n_p$. Then, partial velocities of the 
electrons and positrons can be expressed in terms of ${\bf V}$ 
and ${\bf j}$:
\begin{equation}
{\bf V}_e = \frac{1}{2 n_e} \left( n {\bf V} - \frac{{\bf j}}{e} \right), 
\;\;\;
{\bf V}_p = \frac{1}{2 n_p} \left( n {\bf V} + \frac{{\bf j}}{e} \right).
\end{equation}  
If the number density of plasma, $n$, is much greater than the
charge number density, $|n_p-n_e|$, then $V \gg j/en$. In the general 
case, the hydrodynamic and current velocities can be comparable in the
electron-positron plasma.

MHD equations governing the electron-positron plasma can be obtained 
from the partial momentum equations for the electrons and 
positrons in the standard way (see Urpin 2012). Assuming that plasma 
is non-relativistic, the momentum equation for particles of the sort 
$\alpha$ ($\alpha = e, p$) reads
\begin{eqnarray}
m_{\alpha} n_{\alpha} \left[
\dot{{\bf V}}_{\alpha}
+ ({\bf V}_{\alpha} \cdot \nabla) {\bf V}_{\alpha} \right]
= - \nabla p_{\alpha} 
+ n_{\alpha} {\bf F_{\alpha}} +
\nonumber \\
e_{\alpha} n_{\alpha} \left({\bf E} + \frac{{\bf V}_{\alpha}}{c}
\times {\bf B} \right) + {\bf R}_{\alpha}
\end{eqnarray}  
(see, e.g., Braginskii 1965 where the general plasma formalism
is considered); the dot denotes the partial time 
derivative. Here, ${\bf V}_{\alpha}$ is the mean velocity of 
particles $\alpha$; $n_{\alpha}$ and $p_{\alpha}$ are their 
number density and pressure, respectively; ${\bf F}_{\alpha}$ 
is an external force acting on the particles $\alpha$ (in our 
case ${\bf F}_{\alpha}$ is the gravitational force); ${\bf E}$ 
is the electric field; and ${\bf R}_{\alpha}$ is the internal 
friction force caused by collisions of the particles $\alpha$ 
with other sorts of particles. Since ${\bf R}_{\alpha}$ is the 
internal force, the sum of ${\bf R}_{\alpha}$ over $\alpha$ is 
zero in accordance with Newton's Third Law. Therefore, we have
in the electron-positron plasma ${\bf R}_e = - {\bf R}_p$.

The inertial terms on the l.h.s. of Eq.(3) give a small 
contribution to the force balance because of a small mass of 
both electrons and positrons. Gravitational force can also be 
neglected because of the same reason. A gas pressure is much
smaller than the magnetic pressure in the force-free pulsar 
magnetosphere. Therefore, the momentum equation (3) reads in
the electron-positron plasma
\begin{equation}
e_{\alpha} n_{\alpha} \left({\bf E} + \frac{{\bf V}_{\alpha}}{c}
\times {\bf B} \right) + {\bf R}_{\alpha} = 0.
\end{equation}  
Generally, the friction force, ${\bf R}_{\alpha}$ contains two 
terms: one  proportional to the difference of partial velocities 
$({\bf V}_e - {\bf V}_p)$ and another proportional to the temperature 
gradient (see, e.g., Braginskii 1965). We will neglect the thermal 
contribution to ${\rm R}_{\alpha}$ and take into account only friction 
caused by a difference in the partial velocities. This is equivalent 
to neglecting the thermal diffusion of particles compared to their
hydrodynamic velocities. The friction force is related to the 
velocity difference, $({\bf V}_e - {\bf V}_p)$, by a tensor that 
generally has components along and across the magnetic field and 
the so-called Hall component, which is perpendicular to the both 
magnetic field and velocity difference. In a strong magnetic field, 
parallel and perpendicular components are comparable but the Hall 
component is small (see Braginskii 1965). Therefore, we mimic the 
friction force between electrons and positrons by
\begin{equation}
{\bf R}_e = - \frac{m_e n_e}{\tau_e} ({\bf V}_e - {\bf V}_p),
\end{equation}
where $\tau_e$ is the relaxation time of electrons. Note that 
this simple model for the friction force is often used even for
a magnetized plasma in laboratory conditions (Braginskii 1965) 
and yields qualitatively correct results. We assume that 
accuracy of Eq.(5) is sufficient to study the magnetosphere of 
pulsars. 

It is usually more convenient to use linear combinations of
Eq.(4) than to solve partial equations. The sum of 
electron and positron Eq.(4) yields the equation of hydrostatic 
equilibrium in the magnetosphere 
\begin{eqnarray}
\rho_e {\bf E} + \frac{1}{c} \; {\bf j} \times {\bf B} = 0,
\end{eqnarray} 
where $\rho_e = e (n_p - n_e) = e \delta n$ is the charge density.
Taking the difference between electron and positron Eq.(4),
we obtain the Ohm's law in the form
\begin{equation}
{\bf j} = \rho_e {\bf V} + \sigma \!\left({\bf E} \! + \!
\frac{{\bf V}}{c} \! \times \! {\bf B} \right) 
\end{equation}
where $\sigma = e^2 n_p \tau_e/m_e$ is the conductivity of 
plasma.

It was shown by Urpin (2012) that Eqs.(6)-(7) are equivalent 
to two equations
\begin{equation}
{\bf j} = \rho_e {\bf V}\;, \;\;\;
{\bf E} =  
- \frac{{\bf V}}{c} \times {\bf B}.
\end{equation}
These equations imply that the force-free condition and the
Ohm's law (Eqs.(6)-(7)) are equivalent to the conditions of 
a frozen-in magnetic field and the presence of only advective 
currents in the magnetosphere. Departures from this equivalence
can be caused, for instance, by general relativistic corrections 
(see Palenzuela 2013) but they are very small in the 
magnetosphere. Note that the cross-production of the frozen-in
condition and $\vec{B}$ yields the well-known expression for
the transverse to $\vec{B}$ component of the velocity:  
$\vec{V}_{\perp} = c (\vec{E} \times \vec{B}/B^2)$.
Certainly, the electric current should be 
non-vanishing in the magnetosphere, ${\bf j} \neq 0$, because 
it maintains the magnetic configuration. Hence, the hydrodynamic 
velocity should be non-zero as well since the current is advective. 
Therefore, the force-free magnetosphere can only exist if 
hydrodynamic motions are non-vanishing (Urpin 2012).

\section{Equation governing the magnetospheric waves}

Equation (8) should be complemented by the Maxwell equations. Then, 
the set of equations, governing MHD processes in the force-free 
pulsar magnetosphere reads 
\begin{eqnarray}
\nabla \cdot {\bf E} = 4 \pi \rho_e , \;\;\; \nabla \times
{\bf E} = - \frac{1}{c} \frac{\partial {\bf B}}{\partial t} ,
\nonumber \\
\nabla \cdot {\bf B} = 0 ,\;\;\; 
\nabla \times {\bf B} = \frac{1}{c} 
\frac{\partial {\bf E}}{\partial t} + \frac{4 \pi}{c} {\bf j}, 
\nonumber \\
{\bf j} \approx \rho_e {\bf V}, \;\;\;
{\bf E} \approx  - \frac{{\bf V}}{c} \times {\bf B}.
\end{eqnarray}  

Consider the properties of MHD waves with small amplitudes as 
descibed by these equations. We assune that the electric and 
magnetic fields are equal to ${\bf E}_0$ and ${\bf B}_0$ in 
the unperturbed magnetosphere. The corresponding electric 
current, charge density, and velocity are ${\bf j}_0$, 
$\rho_{e0}$. and ${\bf V}_0$, respectively. For the sake of 
simplicity, we assume that motions in the magnetosphere are 
non-relativistic ($V_0 \ll c$). Linearizing Eq.(9), we obtain 
the set of equations that describes the behaviour of modes 
with a small amplitude. Small perturbations are indicated 
by subscript 1. We consider waves with a short wavelength. 
The space-time dependence of such waves can be taken in the 
form $\propto \exp(i \omega t - i {\bf k} \cdot {\bf r})$, 
where $\omega$ and ${\bf k}$ are the frequency and wave vector, 
respectively. Such waves exist if their wavelength $\lambda = 
2 \pi /k$ is short compared to the characteristic length scale 
of the magnetosphere, $L$. Typically, $L$ is greater than
the stellar radius. We search in magnetohydrodynamic 
modes with the frequency to satisfy the condition $\omega < 
1/ \tau_e$, since we use the MHD approach.  

Substituting the frozen-in condition, ${\bf E} =  - {\bf V}
\times {\bf B}/c$, into the equation $c \nabla \times {\bf E} 
= - \partial {\bf B}/ \partial t$ and linearizing the 
obtained induction equation, we have  
\begin{equation}
i \omega {\bf B}_1 = \nabla \times ( {\bf V}_1 \times {\bf B}_0
+ {\bf V}_0 \times {\bf B}_1 ).
\end{equation}
A destabilising effect of shear already has been studied by Urpin 
(2012). In the present paper, we concentrate on the instability 
caused by the presence of electric currents in the magnetosphere. 
Therefore, we assume that shear is small and neglect terms 
proportional to $|\partial V_{0 i} / \partial x_j|$. Then, Eq.(10) 
reads
\begin{equation}
i\tilde{\omega} {\bf B}_1  = i {\bf B}_0 ({\bf k} \cdot {\bf V}_1 )
- i {\bf V}_1 ({\bf k} \cdot {\bf B}_0 )
- ({\bf V}_1 \cdot \nabla ){\bf B}_0,
\end{equation}
where $\tilde{\omega} = \omega - {\bf k} \cdot {\bf V}_0$. The
last term on the r.h.s. is usually small compared to the third 
term ($\sim \lambda/L$) in a short wavelength approximation. 
However, it becomes crucially important if the wavevector of 
perturbations is almost perpendicular to ${\bf B}_0$.    

Substituting the expression  ${\bf j} = \rho_e {\bf V}$ into
Ampere's law (second line of Eq.(9)) and linearising the 
obtained equation, we have
\begin{equation}
{\bf V}_1 = - \frac{i}{4 \pi \rho_{e0}}  (c {\bf k} \times
{\bf B}_1 + \omega {\bf E}_1) - \frac{\rho_{e1}}{\rho_{e0}} {\bf V}_0 .
\end{equation}
We search in relatively low-frequency magnetohydrodynamic 
modes with the frequency $\omega <  c k$. 
Note that the frequency of MHD modes must satisfy the condition 
$\omega < 1/ \tau_e$ because of the MHD approach used. The
relaxation time can be estimated as $\tau_e \sim \ell_e / c_e$,
where $c_e$ and $\ell_e$ are the thermal velocity and mean free 
path of particles, respectively. The frequency $ck$ can be greater
or smaller than $1/ \tau_e$ depending on a wavelength $\lambda$. 
If $\lambda > 2 \pi \ell_e (c/c_e)$, then we have $ck < 1/ \tau_e$. 
If $\lambda < 2 \pi \ell_e (c/c_e)$, then we have 
$ck > 1/ \tau_e$.  

By eliminating ${\bf E}_1$ from Eq.(12) by making use of the linearised 
frozen-in condition and neglecting terms of the order of $(\omega/ck)
(V_0/c)$, we obtain the following for such modes:
\begin{equation}
{\bf V}_1 + \frac{i \omega}{4 \pi c \rho_{e0}} {\bf B}_0 \times 
{\bf V}_1 = - \frac{i c}{4 \pi \rho_{e0}} {\bf k} \times {\bf B}_1
- \frac{\rho_{e1}}{\rho_{e0}} {\bf V}_0.
\end{equation}
The perturbation of the charge density can be calculated from
the equation $\rho_{e1} = \nabla \cdot {\bf E}_1 /4 \pi$. We have 
then with accuracy in terms of the lowest order in $(\lambda/L)$
\begin{equation}
\rho_{e1} \!=\! \frac{1}{4 \pi c} [ i {\bf B}_0 \cdot 
({\bf k} \!\times\! {\bf V}_1 \!) 
- i {\bf V}_0 \cdot ({\bf k} \!\times\! {\bf B}_1 \!)].
\end{equation}
Substituting Eq.(14) into Eq.(13) and neglecting terms of the
order of $V_{0}^2/c^2$, we obtain the second equation, which
couples ${\bf B}_1$ and ${\bf V}_1$,
\begin{eqnarray}
4 \pi c \rho_{e0} {\bf V}_1 \! + \! i \omega {\bf B}_0 \! \times \!
{\bf V}_1 \! = \!-\! i c^2 {\bf k} \! \times \! {\bf B}_1 \!-\! 
i {\bf V}_0 [ {\bf B}_0 \! \cdot \! ( {\bf k} \! \times \! {\bf V}_1 \! )].
\end{eqnarray}
Eliminating ${\bf B}_1$ from Eqs.(11) and (15) in favor of 
${\bf V}_1$ and again neglecting terms of the order of 
$(\omega/ck)(V_0/c)$ and $(\omega/ck)^2$, we obtain the equation 
for ${\bf V}_1$ in the form 
\begin{equation}
4 \pi \! c \rho_{e0} \! {\bf V}_{1} \!\!-\!\! i \frac{c^2}{\tilde{\omega}}
({\bf k} \! \cdot \! {\bf B}_0) {\bf k} \! \times \!\!{\bf V}_{1} 
\!\!=\! \frac{c^2}{\tilde{\omega}}  
{\bf k} \! \times \! [ (\!{\bf V}_{1} \!\! \cdot \!\! \nabla \!) 
{\bf B}_0
\!-\!i {\bf B}_0 
({\bf k} \! \cdot \!\! {\bf V}_{1}) ]. 
\end{equation} 
It immediately follows from this equation that $({\bf k}
\cdot {\bf V}_1) = 0$ and, hence, the magnetospheric waves 
are transverse. Equation (16) is simplified to
\begin{equation}
\alpha {\bf V}_{1} - i ({\bf k} \cdot {\bf B}_0) {\bf k} \times {\bf V}_{1} 
=   {\bf k} \times ( {\bf V}_{1} \cdot \nabla ) 
{\bf B}_0,
\end{equation}
where
\begin{equation}
\alpha = 4 \pi \! \rho_{e0} \frac{\tilde{\omega}}{c}
\end{equation}
In the case of a uniform magnetic field, Eq.(17) 
reduces to the equation considered by Urpin (2011).

\section{Dispersion equation and instability of magnetospheric modes}

Generally, the stability properties of perturbations are 
complicated even in the local approximation. Equation (17) can be 
transformed to a more convenient form that does not contain a cross 
production of ${\bf k}$ and ${\bf V}_{1}$. Calculating the cross 
production of ${\bf k}$ and Eq.~(17) and taking into account 
${\bf k} \cdot {\bf V}_1 = 0$, we have
\begin{equation}
{\bf k} \times {\bf V}_1 = - \frac{1}{\alpha} \{ i k^2 ({\bf k}
\cdot {\bf B}_0) {\bf V}_1  - ({\bf V}_1 \cdot \nabla) [ {\bf k}
\times ({\bf k}\times {\bf B}_0)] \}.
\end{equation}
Substituting this expression into Eq.~(17), we obtain
\begin{eqnarray}
\left[ \alpha^2 - k^2 ({\bf k} \cdot {\bf B}_0)^2 
\right] {\bf V}_1 = \alpha ({\bf V}_1 \cdot \nabla) {\bf k} \times
{\bf B}_0 +             \nonumber \\
i ({\bf k} \cdot {\bf B}_0) ({\bf V}_1 \cdot \nabla)
[ {\bf k} ({\bf k} \cdot {\bf B}_0) - k^2 {\bf B}_0 ].
\end{eqnarray} 
The magnetospheric waves can exist in the force-free pulsar 
magnetosphere only if the wavevector ${\bf k}$ and the unperturbed 
magnetic field ${\bf B}_0$ are almost (but not exactly) perpendicular
and the scalar production $({\bf k} \cdot {\bf B}_0)$ is small 
but non-vanishing (see Urpin 2011, 2012). The reason for this is clear 
from simple qualitative arguments. The magnetospheric waves are 
transverse (${\bf k} \cdot {\bf V}_1 =0$), and the velocity of plasma 
is perpendicular to the wave vector. However, wave motions across the 
magnetic field in a strong field are suppressed and the 
velocity component along the magnetic field should be much greater 
than in the transverse (see, e.g., Mestel \& Shibata 1994). 
Therefore, the direction of a wavevector ${\bf k}$ should be close 
to the plane perpendicular to ${\bf B}_0$. That is why we treat 
Eq.~(20) only in the case of small $({\bf k} \cdot {\bf B}_0)$.

Consider Eq.~(20) in the neighbourhood of a point, ${\bf r}_0$, 
using local Cartesian coordinates. We assume that the $z$-axis is 
parallel to the local direction of the unperturbed magnetic field 
and the corresponding unit vector is ${\bf b} = {\bf B}_0({\bf r}_0)
/B_0({\bf r}_0)$. The wavevector can be represented as ${\bf k} = 
k_{\parallel} {\bf b} + {\bf k}_{\perp}$, where $k_{\parallel}$ and 
${\bf k}_{\perp}$ are parallel and perpendicular to the magnetic field 
components of ${\bf k}$, respectively. Then, we have from the 
continuity equation 
\begin{equation}
V_{1z} = - \frac{1}{k_{\parallel}} ({\bf k}_{\perp} \cdot {\bf V}_{1 \perp}).
\end{equation}
Since $k_{\perp} \gg k_{\parallel}$, we have $V_{1z} \gg V_{1 \perp}$ and,
hence, $({\bf V}_1 \cdot \nabla) \approx V_{1z} \partial /\partial z$. 
Therefore, the $z$-component of Eq.~(20) yields the following 
dispersion relation
\begin{equation}
\alpha^2 + A \alpha + i D = 0,
\end{equation}
where
\begin{eqnarray}
A= ({\bf k} \times {\bf b}) \cdot \frac{\partial 
{\bf B}_0}{\partial z}\;,   \nonumber \\ 
D = k^2 ({\bf k} \cdot {\bf B}_0) \left[
{\bf b} \cdot \frac{\partial {\bf B}_0}{\partial z} +
i ({\bf k} \cdot {\bf B}_0) \right]. 
\end{eqnarray}
We neglect in $D$ corrections of the order $\sim \lambda / L$
to $k^2 ({\bf k} \cdot {\bf B}_0)^2$. The roots of Eq.(22) 
correspond to two modes with the frequencies given by
\begin{equation}
\alpha_{1, 2} = - \frac{A}{2} \pm \left( \frac{A^2}{4} - i D \right)^{1/2}.
\end{equation}
If the magnetic field is approximtely uniform along the field 
lines, then $\partial {\bf B}_0 / \partial z \approx 0$, and, hence, 
$A \approx 0$ and $D \approx i k^2 ({\bf k} \cdot {\bf B}_0)^2$. 
In this case, the magnetospheric modes are stable and $\alpha_{1,2} 
\approx \pm \sqrt{-iD}$.  The corresponding
frequency is 
\begin{equation}
\tilde{\omega} \approx \pm \frac{ck}{4 \pi \rho_{e0}} 
({\bf k} \cdot {\bf B}_0).
\end{equation}
These waves have been first studied by Urpin (2011). Deriving 
the dispersion Equation (25), it was assumed that $\omega \ll ck$. 
Therefore, the magnetospheric waves can exist only if $({\bf k} 
\cdot {\bf B}_0)$ is small, as was discussed above: $({\bf k} 
\cdot {\bf B}_0) \ll 4 \pi \rho_{e0}$. Sometimes, it is convenient 
to measure the true charge density, $\rho_{e0}$, in units of the 
Goldreich-Julian charge density, $\rho_{GJ}= \Omega B_0 / 2 \pi 
c$, where $\Omega$ is the angular velocity of a pulsar. Then, 
we can suppose $\rho_{e0} = \xi \rho_{GJ}$, where $\xi$ is a 
dimensionless parameter and, hence, the condition $\omega \ll ck$ 
transforms into
\begin{equation}
2 \xi \Omega \gg c |{\bf k} \cdot {\bf b}|.
\end{equation} 
Obviously, this condition can be satisfied only for waves with 
the wavevector almost perpendicular to ${\bf B}_0$. Note,
however, that if ${\bf k}$ is exactly perpendicular to ${\bf B}_0$
the magnetospheric waves do not exist. 

If $\partial {\bf B}_0 / \partial z \neq 0$, the magnetospheric 
waves turn out to be unstable. The instability is especially 
pronounced if $|{\bf k} \cdot {\bf B}_0| < B_0 / L$. In this 
case, the second term in the brackets of Eq.~(24) is smaller 
than the first one and, therefore, the roots are 
\begin{equation}
\alpha_1 \approx - A + i \frac{D}{A} \;, \;\;\; \alpha_2 = - 
i \frac{D}{A}.
\end{equation}
The coefficient $D$ is approximately equal to
\begin{equation}
D \approx k^2 ({\bf k} \cdot {\bf B}_0) \left(
{\bf b} \cdot \frac{\partial {\bf B}_0}{\partial z} \right). 
\end{equation}
The first and second roots of Eq.~(27) correspond to oscillatory 
and non-oscillatory modes, respectively. The occurence of 
instability is determined by the sign of the ratio $D/A$. If this 
ratio is positive for some wavevector ${\bf k}$, then the 
non-oscillatory mode is unstable but the oscillatory one is stable
for such ${\bf k}$. If $D/A < 0$, then the oscillatory 
mode is unstable but the non-oscillatory one is stable for 
corresponding ${\bf k}$. Note, however, that the frequency of 
oscillatory modes often can be very high and $\omega \gg ck$. Our 
consideration does not apply in this case. Indeed, we have 
$\alpha_1 \sim A$ and, hence, $\tilde{\omega}_1 \sim ck (B_0 /
4 \pi \rho_{e0} L)$. The condition $\omega \ll ck$ implies that 
$B_0/4 \pi \rho_{e0} L <1$. Expressing the charge density in units 
of the Goldreich-Julian density, $\rho_{e0} = \xi \rho_{GJ}$, we 
transform this inequality into
\begin{equation}
\frac{1}{2 \xi} \; \frac{c}{\Omega L} \ll 1.
\end{equation}    
This condition can be satisfied only in regions where $\xi \gg 1$
and the charge density is much greater than the Goldreich-Julian
density. If inequality (29) is not fulfilled, then Eq.~(27) for the 
oscillatory mode $\alpha_1$ does not apply, and only the non-oscillatory 
modes exist. For example, the charge density is large in the region where 
the electron-positron plasma is created. Therefore, condition (29) 
can be satisfied there, and, hence, the oscillatory instability 
can occur in this region. 

The non-oscillatory modes have a lower growth rate and usually can 
occur in the pulsar magnetosphere. For any magnetic configuration, 
it is easy to verify that one can choose the wavevector of 
perturbations, ${\bf k}$, in such a way that the ratio $D/A$ becomes 
positive, and, hence, the non-oscillatory mode is unstable. Indeed, 
we can represent ${\bf k}$ as the sum of components parallel and 
perpendicular to the magnetic field, ${\bf k} = {\bf k}_{\parallel} + 
{\bf k}_{\perp}$. Obviously, $A \propto k_{\perp}$ and $D \propto
k_{\parallel}$ and, hence, $A/D \propto k_{\parallel}/ k_{\perp}$.
Therefore, if $A/D < 0$ for a value of ${\bf k} = ( k_{\parallel},
k_{\perp})$, this ratio changes the sign for ${\bf k} = ( - k_{\parallel},
k_{\perp})$ and ${\bf k} = ( k_{\parallel},- k_{\perp})$, and waves 
with such the wavevectors are unstable. It turns out that there 
always exists the range of wavevectors for which the non-oscillatory 
modes are unstable and, hence, the force-free magnetosphere is 
always the subject of instability.  

The necessary condition of instability is  $D \neq 0$. As it was 
mentioned, the magnetospheric waves exist only if the wavevector 
${\bf k}$ is close to the plane perpendicular to the unperturbed 
magnetic field, ${\bf B}_0$, and the scalar production $({\bf k} \cdot 
{\bf B}_0)$ is small (but non-vanishing). Therefore, the necessary
condition $D \neq 0$ is equivalent to ${\bf b} \cdot (\partial {\bf B}_0/
\partial z) \neq 0$. Since ${\bf b} = {\bf B}_0 / B_0$, we can rewrite
this condition as
\begin{equation}
{\bf B}_0 \cdot \frac{\partial {\bf B}_0}{\partial z} \neq 0.
\end{equation}  
This condition is satisfied if the magnetic pressure gradient along 
the magnetic field is non-zero. The topology of the magnetic field 
can be fairly complicated in the magnetosphere, particularly in a 
region close to the neutron star. This may happen because the field 
geometry at the neutron star surface should be very complex (see, 
e.g., Bonanno et al. 2005, 2006). Therefore, Eq.~(30) can be 
satisfied in different regions of the magnetosphere.
 However, this condition can be fulfilled even if the magnetic
configuration is relatively simple. As a possible example, we 
consider a region near the magnetic 
pole of a neutron star. It was shown by Urpin (2012) that a 
special type of cylindrical waves can exist there
with $m \neq 0$, where  $m$ is the azimuthal wavenumber. The
criterion of instability (30) is certainly satisfied in this region
and, hence, the instability can occur. Indeed, one can mimic the 
magnetic field by a vacuum dipole near the axis. The radial and 
polar components of the dipole field in the spherical 
coordinates $(r, \theta, \varphi)$ are
\begin{equation}
B_r = B_p \left( \frac{a}{r} \right)^3 \cos \theta , \;\;\;
B_{\theta} = \frac{1}{2} B_p \left( \frac{a}{r} \right)^3 \sin \theta ,
\end{equation} 
where $B_p$ is the polar strength of the magnetic field at the 
neutron star surface and $a$ is the stellar radius (see, e.g., 
Urpin et al. 1994). The radial field is much greater than the 
polar one near the axis and, therefore, it is easy to check that 
the criterion of instability (30) is fulfilled in the polar gap. 
Hence, filament-like structures can be formed there. In some 
models, note that the force-free field at the top of the polar gap 
can differ from that of a vacuum dipole (see, e.g., Petrova 2012) 
but this cannot change our conclusion qualitatively. We will
consider the instability in the polar gap in more detail 
elsewhere.  

From Eq.~(30), it follows that the instability in 
pulsar magnetospheres is driven by a non-uniformity of the magnetic 
pressure and, hence, it can be called ``the magnetic pressure-driven 
instability''. Note that this instability can occur only in plasma 
with a non-zero charge density, $\rho_{e0} \neq 0$, and does not 
arise in a neutral plasma with $\rho_{e0} = 0$.

\section{Discussion}

We have considered stability of the electron-positron plasma
in the magnetosphere of pulsars. The pair plasma in the 
magnetosphere is likely created in a two-stage process: primary 
particles are accelerated by an electric field parallel to the 
magnetic field near the poles up to extremely high energy, and 
these produce a secondary, denser pair plasma via a cascade process 
(see, e.g., Michel 1982). The number density of this secondary 
plasma greatly exceeds the Goldreich-Julian number density, $n_{GJ} 
= |\rho_{GJ}|/e$, required to maintain a corotation electric field 
and, hence, the multiplicity factor $\xi$ can be very large. 
Unfortunately, this factor is model dependent and rather 
uncertain with estimates in a wide range from $10^2$ to $10^6$
(see, e.g., Gedalin et al. 1998). For example, the model of 
bound-pair creation above polar caps (Usov \& Melrose 1996) 
results in a relatively low value of $\xi$. This model 
postulates that the photons emitted by primary particles as a 
result of curvature emission create bound pairs (positronium)
rather than free pairs. While the pairs remain bound, the 
screening of the component ${\bf E}$ parallel to the magnetic 
field is inefficient because screening is attributed to free 
pairs that can be charge-separated as a result of acceleration 
by ${\bf E}_{\parallel}$. In the absence of screening, the height 
of the polar gap and the maximum energy of primary particles 
increase. The main part of the energy that primary particles 
gain during their motion through the polar gap is transformed 
into the energy of curvature photons and then into the energy
of secondary pairs. Assuming that the electric field in the
gap has more or less standard value ($\sim 10^{10}$ V cm$^{-1}$),
Usov \& Melrose (1996) obtain the following estimate for the
multiplicity factor above polar caps:
\begin{equation}
\xi \sim 4 \times 10^2 \left( \frac{P}{0.1 {\rm s}} \right)^{-3/4},
\end{equation}  
where $P$ is the pulsar period. Note that this value can be 
higher if dissociation of bound pairs is taken into account. 
However, the instability can operate in the region of pair
creation even if $\xi$ is such small. It can be efficient in
regions with $\xi \sim 1$ as well.

The geometry of motions in the unstable magnetospheric waves is 
simple. Since these waves are transverse (${\bf k} \cdot 
{\bf V}_1 = 0$) and the wavevector of such waves should be 
close to the plane perpendicular to ${\bf B}_0$, plasma motions
are almost parallel or anti-parallel to the magnetic field.
The velocity across ${\bf B}_0$ is small. In our model, 
we have only considered the stability of plane waves using a 
local approximation. In this model, the instability  
should lead to formation of filament-like structures with filaments 
alongside the magnetic field lines. Note that plasma can move 
in the opposite directions in different filaments. The 
characteristic timescale of formation of such structures is 
$\sim 1/ {\rm Im} \omega$. Since the necessary condition (30) 
is likely satisfied in a major fraction of a magnetosphere, 
one can expect that filament-like structures can appear (and 
disappear) in different magnetospheric regions. We used the 
hydrodynamic approach in our consideration, which certainly does not 
apply to a large distance from the pulsar where the number
density of plasma is small. Therefore, the considered instability
is most likely efficient in the inner part of a magnetosphere
where filament-like structures can be especially pronounced.
The example of a region where the instability can occur is the 
so-called dead zone. Most likely, the hydrodynamic approximation
is valid in this region and hydrodynamic motions are non-relativistic,
as it was assumed in our analysis. Note that a particular geometry
of motions in the basic (unperturbed) state is not crucial for the
instability and cannot suppress the formation of 
filament-like structures. These structures can be responsible for 
fluctuations of plasma and, hence, the magnetospheric emission can 
fluctuate with the same characteristic timescale.

It should be also noted that the considered instability
is basically electromagnetic in origin as followed in
our treatment. Hydrodynamic motions in the basic state play no 
important role in the instability. For instance, the unperturbed
velocity does even not enter the expression for the growth rate.
Therefore, one can expect that the same type of instability 
arises in the regions where velocities are relativistic. This 
case will be considered in detail elsewhere.

Hydrodynamic motions accompanying the instability can be 
the reason of turbulent diffusion in the magnetosphere.
This diffusion should be highly anisotropic because both 
the criteria of instability and its growth rate are sensitive 
to the direction of the wave vector. However, the turbulent 
diffusion caused by motions may be efficient in the transport of 
angular momentum and mixing plasma with a much higher 
enhancement of the diffusion coefficient in the direction
of the magnetic field since the velocity of motions across the 
field is much slower than along it. 

The characteristic growth rate of unstable waves, Im~$\omega$,  
can be estimated from Eq.~(27) as Im~$\omega \sim (c/ 4\pi 
\rho_{e0})(D/A)$. Since ${\bf k}$ and ${\bf B}_0$ should be 
close to orthogonality in magnetospheric waves, we have 
$A \sim k B_0 / L$ and $D \sim k^2 ({\bf k} \cdot {\bf b}) 
B_0^2 /L$, where we estimate ${\bf b} \cdot ( \partial {\bf B}_0
/\partial z)$ as $B_0/L$. Then,
\begin{equation}
{\rm Im}~\omega \sim c k ~ \frac{({\bf k} \cdot {\bf B}_0)}{4
\pi \rho_{e0}}
\sim  \frac{1}{\xi} c k ~ \frac{c({\bf k} 
\cdot {\bf b})}{\Omega}.
\end{equation} 
Like stable magnetospheric modes, the unstable ones can 
occur in the magnetosphere if Eq.~(26) is satisfied. Generally,
this condition requires a position close to orthogonality of
(but not orthogonal) ${\bf k}$ and
${\bf B}_0$. Under certain conditions, however, the instability 
can arise even if departures from orthogonality are not very 
small but $\xi \gg 1$. As it was mentioned, this can happen in 
regions where the electron-positron plasma is created. The 
parameter $\xi$ can also be greater than 1  in those regions 
where plasma moves with the velocity greater $\Omega L$. Indeed, 
we have $\rho_{e0} = (1/4 \pi) \nabla \cdot {\bf E}_0$ for the 
unperturbed charge density. Since ${\bf E}_0$ is determined by the 
frozen-in condition (8), we obtain $\rho_{e0} \sim (1/4 \pi c L) 
V_0 B_0$. If the velocity of plasma in a magnetosphere is greater 
than the rotation velocity, then $\xi \sim V_0 / \Omega L$. Some
models predict that the velocity in the magnetisphere can reach
a fraction of $c$. Obviously, in such regions, condition (26) 
is satisfied even if departures from orthogonality of ${\bf k}$
and ${\bf B}_0$ are relatively large.  

The growth rate of instability (31) is sufficiently high and 
can reach a fraction of $ck$. For example, if a pulsar rotates 
with the period 0.01 sec and $\xi \sim 1$, magnetospheric waves 
with the wavelength $\sim 10^5 - 10^6$ cm grow on a timescale 
$\sim 10^{-4} - 10^{-5}$ s if a departure from orthogonality 
between ${\bf k}$ and ${\bf B}_0$ is of the order of $10^{-4}$.  
The considered instability can occur everywhere in the
magnetosphere except regions close to the surfaces where
${\bf B}_0 \cdot (\partial {\bf B}_0 /\partial z) = 0$ and 
instability criterion (30) is not satisfied.

The instability considered is caused by a combined 
action of non-uniform magnetic field and non-zero charge
density. Certainly, this is not the only instability that
can occur in the pulsar magnetosphere. There are many
factors that can destabilize a highly magnetized 
magnetospheric plasma and lead to various instabilities
with substantially different properties. For instance, 
differential rotation predicted by many models of the 
magnetosphere can be the reason of instability as was 
shown by Urpin (2012). This instability is closely related 
to the magnetorotational instability (Velikhov 1959) which is 
well-studied in the context of accretion disks (see, e.g., 
Balbus \& Hawley 1991). Generally, the regions, where rotation 
is differential and the magnetic field is non-unifom, can 
overlap. Thus, the criteria of both instabilities can be 
fulfilled in the same region. However, these instabilities 
usually have substantially different growth rates. The 
instability caused by differential rotation arises usually 
on a time-scale comparable to the rotation period of a 
pulsar. The growth rate of the magnetic pressure-driven 
instability is given by Eq.(30) and can even reach a fraction 
of $ck$ in accordance with our results. Therefore, this 
instability occurs typically on a shorter time-scale than the 
instability caused by differential rotation. If two 
different instabilities can occur in the same region, then, 
the instability with a shorter growth time usually turns 
out to be more efficient and determines plasma fluctuations.
It is likely, therefore, that the magnetic pressure-driven
instability is more efficient everywhere in the
magnetosphere except surfaces where criterion (30) is
not satisfied. In the neighbourhood of these surfaces,
howevere, the instability associated with differential
rotation can occur despite it arises on a longer time-scale.
Therefore, it appears that the whole pulsar magnetosphere 
should be unstable.

Instabilities can lead to a short-term variability
of plasma and, hence, to modulate the magnetospheric emission of pulsars. 
The unstable plasma can also modulate the radiation produced at the 
stellar surface and propagating through the magnetosphere.
Since the growth time of magnetospheric waves can be substantially
different in different regions, the instability can lead to a 
generation of fluctuations over a wide range of timescales,
including those yet to be detected in the present and future
pulsar searches (Liu et al. 2011, Stappers et al. 2011).
Detection of such fluctuations would uncover the physical 
conditions in the magnetosphere and enable one to construct
a relevant model of the pulsar magnetosphere and its observational
manifestations beyond the framework of the classical concept
(see, e.g., Kaspi 2010).


\section*{Acknowledgement}

The author thanks the Russian Academy of Sciencs for financial support 
under the programm OFN-15.


\begin{thebibliography}{}

\bibitem{}
Balbus S., Hawley J. 1991. ApJ, 376, 214

\bibitem{}
Beskin V. 1997. Uspekhi Fizicheskikh Nauk, 167, 689

\bibitem{}
Bonanno A., Urpin V. 2008a. A\&A, 477, 35

\bibitem{}
Bonanno A., Urpin V. 2008b. A\&A, 488, 1

\bibitem{}
Bonanno A., Urpin V, Belvedere G. 2005. A\&A, 440, 199

\bibitem{}
Bonanno A., Urpin V, Belvedere G. 2006. A\&A, 451, 1049

\bibitem{}
Braginskii S. 1965. Review of Plasma Physics, 1, 205

\bibitem{}
Contopoulos I., Kazanas D., \& Fendt C. 1999. ApJ, 511, 351

\bibitem{}
Davidson R. 1990. Physics of non-neutral plasmas ( Addison-Wesley
Publishing Company)

\bibitem{}
Davidson R., Felice G. 1998. PhPl, 5, 3497

\bibitem{}
Fung P.K., Khechinashvili D., \& Kuijpers J. 2006. A\&A, 445, 779

\bibitem{}
Gedalin M., Melrose D., \& Gruman E. 1998. PRE, 57, 3399


\bibitem{}
Goodwin S., Mestel J., Mestel L., \& Wright G. 2004. MNRAS, 
349, 213


\bibitem{}
Kaspi V.M. 2010. Publications of the National Academy of
Science, 107, No. 16, 7147  

\bibitem{}
Komissarov S. 2006. MNRAS,367, 19

\bibitem{}
Levy R. 1965. PhPl, 8, 1288

\bibitem{}
Liu  K. et al. 2011. MNRAS, 417, 2916

\bibitem{}
McKinney J.C. 2006. MNRAS, 368, L30

\bibitem{}
Mestel L. 1973. Ap\&SS, 24, 289

\bibitem{}
Mestel L., Shibata S. 1994. MNRAS, 271, 621

\bibitem{}
Michel F. 1973. ApJ, 180, L133

\bibitem{}
Michel F. 1982. Rev. Mod. Phys., 54, 1



\bibitem{}
Palenzuela C. 2013. MNRAS (in press)

\bibitem{}
Petri J., Heyvaerts J., \& Bonazzola S. 2002. A\&A, 287, 520

\bibitem{}
Petri J., Heyvaerts J., \& Bonazzola S. 2003. A\&A, 411, 203

\bibitem{}
Petrova S. 2013. ApJ, 764, 129

\bibitem{}
Petrova S. 2012. MNRAS, 427, 514

\bibitem{}
Stappers B. et al. 2011. A\&A, 530, A80


\bibitem{}
Tayler R. 1973a. MNRAS, 161, 365

\bibitem{}
Tayler R. 1973b. MNRAS, 163, 77 

\bibitem{}
Timokhin A. 2006. MNRAS, 368, 1055

\bibitem{}
Urpin V. 2011. A\&A, 535, L5

\bibitem{}
Urpin V. 2012. A\&A, 541, 117

\bibitem{}
Urpin V., Chanmugam G., \& Sang Y. 1994. ApJ, 433, 780 

\bibitem{}
Velikhov E. 1959. Sov. Phys. JETP, 9, 995 



















\end{thebibliography}
\end{document}